\documentclass[doublecol,figures]{epl2} 

\title{Nonequilibrium dynamics of an exactly solvable Ising--like
  model and protein translocation}
\shorttitle{Exactly solvable model of protein translocation} 

\author{A. Pelizzola\inst{1,2,3}\thanks{E-mail: \email{alessandro.pelizzola@polito.it}} \and M. Zamparo\inst{3}\thanks{E-mail: \email{marco.zamparo@hugef-torino.org}}}
\shortauthor{A. Pelizzola \etal}

\institute{                    
  \inst{1} Dipartimento di Scienza Applicata e Tecnologia, CNISM and
  Center for Computational Studies, Politecnico di Torino,
Corso Duca degli Abruzzi 24, I--10129 Torino, Italy\\
  \inst{2} INFN, Sezione di Torino, via Pietro Giuria 1, I-10125 Torino,
Italy\\
\inst{3} Human Genetics Foundation, HuGeF, Via Nizza 52, I-10126 Torino,
Italy
}


\abstract{ Using an Ising--like model of protein mechanical unfolding,
  we introduce a diffusive dynamics on its exactly known free energy
  profile, reducing the nonequilibrium dynamics of the model to a
  biased random walk. As an illustration, the model is then applied to
  the protein translocation phenomenon, taking inspiration from a
  recent experiment on the green fluorescent protein pulled by a
  molecular motor. The average translocation time is evaluated
  exactly, and the analysis of single trajectories shows that
  translocation proceeds through an intermediate state, similar to
  that observed in the experiment.  }

\begin{document}

\maketitle

\section{Introduction}

In an attempt to understand the physics of biomolecules, statistical
physicists have developed models at various levels of
coarse--graining, from all--atom models down to simple Ising--like
models. Classical examples in the latter category are the Zimm--Bragg
model of the helix--coil transition \cite{ZimmBragg} and the Poland--Scheraga
model of DNA denaturation \cite{PolandScheraga}.

In the effort of developing models for protein folding, several
Ising--like models have been proposed
\cite{WS1,WS2,ME1,ME2,Finkelstein,Baker}. One of these,
sometimes called Wako--Sait\^{o}--Mu\~{n}oz--Eaton (WSME) model
\cite{WS1,WS2,ME1,ME2,ME3,Maritan}, has recently been the subject of
some research activity, since its thermodynamics is exactly solvable
\cite{BP-PRL02,MioJSTAT,BPZ-PRL07,MarcoJSTAT}. In one research line,
the model was generalized to describe mechanical unfolding
\cite{IPZ-PRL07,IPZ-JCP07,IP-PRL08,IPZ-PRL09,CaraglioFnIII,CIP-PRE11,Samori},
and its nonequilibrium kinetics was studied through Monte Carlo
simulations.

Here we take a different approach, which does not make use of Monte
Carlo simulations: exploiting the mathematical properties of the model
which allow an exact numerical computation of its free energy profile
as a function of a suitable reaction coordinate, we define a diffusive
dynamics on such free energy profile and reduce the nonequilibrium
kinetics of the model to a biased one--dimensional random walk on the
chosen reaction coordinate. This allows us on the one hand to exactly
calculate quantities like mean first passage times, and on the other
hand to easily simulate single trajectories.

\begin{figure}
\includegraphics*[width=0.48\textwidth]{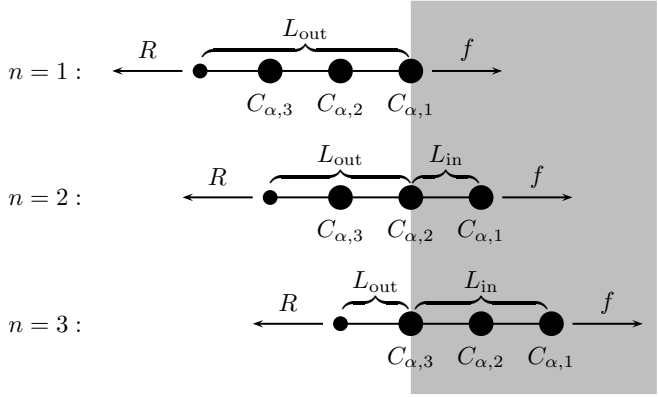}
\caption{Illustration of our setup for $N = 3$ and $n = 1, 2,
  3$. The portion of the chain outside the pore is represented, for
  simplicity only, in a fully extended configuration. \label{fig:illu}}
\end{figure}

As an illustration of our approach, we study the translocation of a
protein through a narrow, long and neutral pore, under the action of
an importing and a resisting force.  Protein translocation is a
nonequilibrium phenomenon, which in the recent years, also thanks to
the development of single--molecule techniques, has been the subject
of an intense research activity, both experimental
\cite{Fernandez-Cell03,Kenniston-PNAS05,Sauer-NSMB08,Lang-Cell11,Bustamante-Cell11}
and theoretical
\cite{Makarov-JCP05,Makarov-PSSB06,Cecconi-JPCB09,Cecconi-PM11,Stan-PNAS11}.
In our model, the pore is described in a very simplified way, by
imposing the constraint that the aminoacids imported in the pore must
be in an unfolded state, and the natural reaction coordinate is the
number of imported aminoacids. Our setup, involving an importing and a
resisting force, is illustrated in Fig.\ \ref{fig:illu}, and is inspired by a
recent experiment \cite{Bustamante-Cell11}.

In \cite{Bustamante-Cell11}, translocation of the green
fluorescent protein (GFP) by a molecular motor was studied and
quantitatively characterized by means of a single molecule
experiment. Using suitable handles, the C--terminal of GFP was
attached to the molecular motor ClpX (alone, or bound to the peptidase
ClpP which degrades the translocated protein), while a resisting force
was applied to the N--terminal by means of a suitable optical trap
apparatus. ClpX was shown to exert a mechanical force to unfold GFP
and translocate it in a stepwise fashion through its pore. Among other
results, the translocation velocity was evaluated as a function of the
resisting force, and it was shown that GFP unfolding is most often a
2--stage phenomenon, proceeding through an intermediate state.

In our study we shall focus on the above two aspects of this
phenomenon, which can be easily described in the framework of our
simple model: by studying the mean translocation time we shall show
that it depends non--monotonically on the resisting force, while by
simulating single trajectories we show that unfolding occurs through
an intermediate state, which has the same structure observed in the
experiment.

\section{Model}

The model has 2 sets of binary degrees of freedom. For a protein of
$N$ aminoacids, we associate a variable $m_k$ to each aminoacid $k =
1, \ldots N$, taking values $0, 1$. $m_k = 1 (0)$ represents an
aminoacid in a native--like (unfolded) configuration. Two aminoacids
$i$ and $j > i$ can interact only if they are in contact in the native
state and all aminoacids from $i$ to $j$ are native--like. Given a
configuration $m = \{ m_k \}$, an additional set $\sigma = \{
\sigma_{ij} \}$ of binary variables is defined, which specifies the
orientation of the protein chain with respect to an external
force. More precisely, we associate an orientational variable
$\sigma_{ij} = \pm 1$ to each portion of the chain delimited by two
non--native aminoacids $i$ and $j > i$, such that $(1 - m_i) (1 - m_j)
\prod_{k=i+1}^{j-1} m_k = 1$. Given a reference direction (which in
the following will be the direction of the forces acting on the
protein), $\sigma_{ij} = +1$ (-1) means that the stretch from $i$ to
$j$ is parallel (antiparallel) to the reference
direction. Pictorially, we can think to the protein backbone as
represented by the sequence of $C_\alpha$ atoms, divided into
native--like stretches (which can be as short as the link between two
consecutive $C_\alpha$'s and as long as the whole chain) that can
rotate around the ``unfolded'' $C_\alpha$'s. Everything is then
reduced to a 1--dimensional projection along the reference direction
and the end--to--end lengths $\{l_{ij}\}$ of the native stretches are
read from the Protein Data Bank (PDB).

In the present application to the translocation problem we consider a
narrow, long and neutral pore, and introduce an additional degree of
freedom, $n$, which specifies the position of the protein with respect
to the pore. The portion of the chain from aminoacid 1 to $n$ is
inside the pore, frozen in an unfolded, extended state, represented by
the conditions $m_k = 0\ (k = 1, \ldots n)$ and $\sigma_{k,k+1} = +1\
(k = 1, \ldots n-1)$. Physically, the pore is assumed to be (i)
narrow, so that the protein must unfold and orient in the force
direction to enter, (ii) sufficiently long to contain the whole
protein, so that refolding is not possible after translocation (one
might as well think that the protein has been degraded, as it can
happen with the molecular motors studied in \cite{Bustamante-Cell11}),
and (iii) neutral (no interaction with the protein except for the
above--mentioned geometrical constraints). The remaining portion, from
aminoacid $n+1$ to $N$, is outside the pore and its degrees of freedom
can vary as described above.

The model can be defined through the following Hamiltonian:

\begin{equation}
\label{eq:H}
H(m,\sigma,n) = - \epsilon \sum_{i=n+1}^{N-1} \sum_{j=i+1}^N \Delta_{ij}
\prod_{k=i}^j m_k - f L_\tx{in} + R L_\tx{out}.
\end{equation}

Here $\Delta$ is a contact matrix: its element $\Delta_{ij}$ is
defined as the number of atomic contacts between aminoacids $i$ and
$j$, where we have an atomic contact whenever 2 atoms (hydrogens
excluded) are closer than 4 \AA\ in the native configuration reported
in the PDB. $\epsilon$ is the contact interaction
energy, and all energies will be defined in units of $k_B T$, where
$k_B$ is Boltzmann's constant and $T$ absolute temperature. In the
following we choose $\epsilon = 0.13$, a value at which GFP (PDB code
1B9C, chain A) in absence of forces is fully native (the average
fraction of native contacts is $Q = 0.9827$, while the midpoint of the
denaturation transition, where $Q = 1/2$, corresponds to $\epsilon =
0.116$).

The model contains also 2 force terms. $f$ is the importing force,
exerted by the molecular motor. In the picture of a long pore, we can
think that this force is applied to aminoacid number 1. Assuming that
the pore position is fixed, this force is coupled to $L_\tx{in} =
\sum_{i=1}^{n-1} l_{i,i+1}$, the length of the portion of the chain
inside the pore. According to the so--called power--stroke scenario
\cite{Lang-Cell11,Bustamante-Cell11}, the actual force generated by
the molecular motor is believed to be a time--dependent force, made of
short pulses, and we can think that $f$ is the corresponding time
average. $R$ is the resisting force, which in \cite{Bustamante-Cell11}
was exerted by the optical tweezers apparatus. This is coupled to the
length of the portion of the chain which is outside the pore,
\begin{equation}
\label{eq:Lout}
L_\tx{out} = -
\sum_{i=n}^{N} \sum_{j=i+1}^{N+1} l_{ij} \sigma_{ij} (1 - m_i) (1 - m_j)
\prod_{k=i+1}^{j-1} m_k, 
\end{equation}
with the boundary condition $m_{N+1} = 0$ (an always unfolded variable
is associated to the last atom of aminoacid number $N$, to which $R$
is applied). In the following we shall assume that $R$ is kept
constant during the unfolding and translocation processes: several
measurements were taken in \cite{Bustamante-Cell11} under this
condition, which can be experimentally realized by means of a suitable
feedback system. 

As argued in \cite{Lang-Cell11}, the conformational degrees of
freedom, here represented by $m$ and $\sigma$, should have a much
faster dynamics than the translational degree of freedom, here $n$. We
therefore sum over $m$ and $\sigma$, obtaining the effective free
energy
\begin{equation}
\label{eq:g}
g(n) = - \ln \sum_{m,\sigma} \exp[-H(m,\sigma,n)],
\end{equation}
where the sum can be evaluated in a numerically exact way by means of
a polynomial (in $N$) recursive algorithm \cite{IPZ-JCP07}. 

We use the above free energy to define a driven--diffusive dynamics
for our model. At each time step, the number $n$ of imported
aminoacids can vary by $\pm 1$ or 0. We have considered several
choices for the transition probability, namely heat bath
\begin{equation}
\label{eq:HB}
W(n \to n + \Delta n) = \frac{\exp[-g(n + \Delta n)]}
{\sum_{\Delta n' = \pm 1, 0} \exp[-g(n + \Delta n')]},
\end{equation}
which does not satisfy detailed balance, and 2 choices satisfying
detailed balance, Metropolis and Glauber, where
\begin{eqnarray}
\label{eq:Metro}
W(n \to n \pm 1) &=& \frac{1}{2} w(g(n \pm 1) - g(n)),
\nonumber \\
 W(n \to n) &=& 1 - \sum_{\Delta n' = \pm 1} W(n \to n + \Delta n')
\end{eqnarray}
and $w(\Delta g) = \min ( 1, \exp(-\Delta g) )$ for Metropolis, 
$w(\Delta g) = [ 1 + \exp(\Delta g) ]^{-1}$ for Glauber.
The above rules are supplemented by suitable boundary conditions. We
have an absorbing boundary at $n = N$, with $W(N \to N) = 1$, $W(N \to
N \pm 1) = 0$, meaning that the translocation process is considered
completed when all aminoacids have been imported into the pore. The
boundary at $n = 1$ is instead partially reflecting: in Eq.\
\ref{eq:HB} $\Delta n = -1$ is forbidden and the normalization is
modified accordingly, with similar changes for the Metropolis and
Glauber choices. This means that in our model the protein cannot
detach from the molecular motor. 


Since we assume protein degrees of freedom are equilibrated
  during translocation events, our model should be more properly
  understood on a mesoscopic level rather than on a microscopic scale,
  the only one where dynamics is {\it a priori} expect to be
  reversible. This fact has stimulated us to consider a possible
  violation of detailed balance (heat bath) and to compare the
  corresponding kinetics with reversible evolutions (Metropolis and
  Glauber). However, this comparison has revealed that there is not
  any qualitative difference among such prescriptions within the
  present model, apart from the fact that heat bath is (roughly twice)
  faster than Metropolis, which in turn is faster than Glauber. As a
  consequence, the results reported below, which have been obtained
  with the heat bath dynamics, are rather robust.

Irrespectively of the choice of the
dynamics, the evolution of the number $n$ of imported aminoacids is
described by a simple stochastic process: a biased random walk on a
lattice segment, with a partially reflecting and an absorbing
boundary. Setting $n = 1$ as our initial condition at time $t = 0$, we
define the translocation time $T$ as the first passage time at $n =
N$: its average value will therefore be a mean first passage time,
which can be evaluated by the generating function method
\cite{Feller}, obtaining $\langle T \rangle = \sum_{n=1}^{N-1} \langle
t \rangle_n$, where the average time spent at position $n$ is
\begin{equation}
\langle t \rangle_n = \frac{1}{W(n \to n+1)} \sum_{m=n}^{N-1} 
\prod_{l=m+1}^n \frac{W(l \to l-1)}{W(l \to l+1)}.
\end{equation}

\section{Results}

\begin{figure}
\includegraphics*[width=0.48\textwidth]{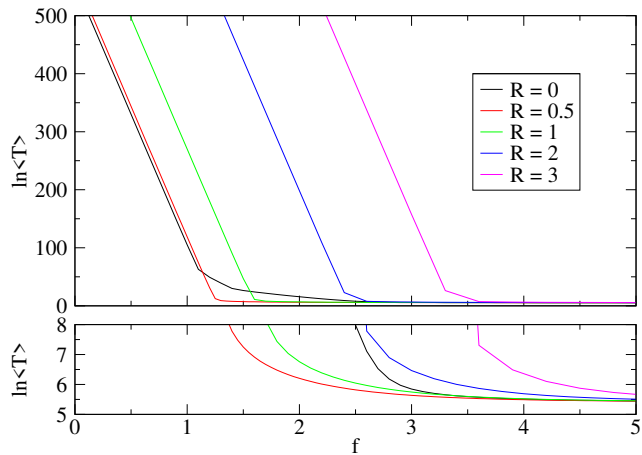}
\caption{
Logarithm of the translocation time $\langle T\rangle$ as a function of importing
  force and blow-up behavior at low $\ln \langle T\rangle$.
\label{fig:t_vs_f}}
\end{figure}

We have evaluated the average translocation time $\langle T\rangle$ of
GFP as a function of the importing force $f$, applied to the
C--terminal, and the resisting force $R$, applied to the
N--terminal. In Fig.\ \ref{fig:t_vs_f} we report $\ln \langle
T\rangle$ as a function of $f$ for several values of the resisting
force $R$.  In order to understand the 
content of the figure and the following discussion it is
important to observe that both the time needed to unfold the molecule
and the time needed to enter the pore contribute to $\langle
T\rangle$.

Two regimes are clearly observed. For small importing force, $\ln
\langle T\rangle$ decreases linearly with $f$, consistently with
Bell's theory \cite{Bell,KumarLi} applied to mechanical unfolding in
presence of a barrier. Indeed, in this regime, translocation is
extremely slow because the importing force is not sufficient to
overcome the free energy barrier to unfolding. As a consequence,
$\langle T\rangle$ is dominated by the unfolding time.

For large importing force, $\ln \langle T\rangle$ is, on the scale of
this graph, roughly independent on $f$. In this regime $f$ can easily
unfold the protein, which can then enter the pore. However, as we
shall see later, the average translocation time still depends on $f$
and $R$. It can be seen that for some values of the resisting force
$R$ the transition between the 2 regimes occurs in 2 steps: this is
the first signature we encounter of the presence of intermediates in
GFP translocation (this is expected, since it is known that
intermediates are found in GFP mechanical unfolding
\cite{Rief-PNAS04,Rief-PNAS07}).



%

\begin{figure}
\includegraphics*[width=0.48\textwidth]{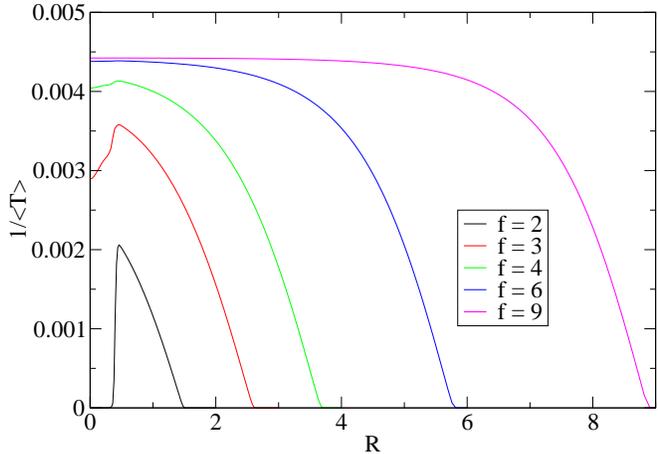}
\caption{Inverse translocation time as a function of resisting force. \label{fig:t_vs_R}}
\end{figure}

In Fig.\ \ref{fig:t_vs_R} we report $1/\langle T\rangle$ as a function
of the resisting force $R$ for different values of the importing force
$f$. Several interesting phenomena can be observed. First of all, let
us consider the $f = 2$ curve. Even without resisting force, $R = 0$,
this importing force is too weak to unfold the protein and the inverse
translocation time $1/\langle T\rangle$ (proportional to the
translocation velocity) is practically zero on the scale of the graph:
more precisely, this corresponds to the linearly decreasing branch in
Fig.\ \ref{fig:t_vs_f}. Upon increasing $R$ we see that its effect is
non--trivial: a small resisting force helps unfolding the protein, and
after the unfolding the translocation becomes possible. Finally, by
further increasing $R$, the translocation velocity decreases and
eventually vanishes again. In this latter regime the protein is fully
extended by the joint action of $f$ and $R$, and the translocation
velocity is determined by the competition of these 2 forces.

Increasing the importing force to $f = 3$ we see that the
translocation velocity is nonzero already at $R = 0$ and for small
increasing $R$ it grows up to a maximum. In this portion of the curve
we can distinguish 2 regimes, separated by a change in slope. This is
another signature of the presence of an intermediate: a small
resisting force is more likely to unfold our protein only partially,
while a larger $R$ can more easily give rise to complete unfolding.

The curve at $f = 4$ can be qualitatively compared to the experimental
results for the translocation velocity (which was measured in
\cite{Bustamante-Cell11}, see Fig.\ 2A, as a function of the resisting
force): both exhibit a small increase at small $R$ and then a more or
less sharp decrease at large $R$. Notice that our model cannot
describe conformational fluctuations inside the pore, so we can
compare, at least qualitatively, our results with the extension
velocity in \cite{Bustamante-Cell11}, but we cannot give an estimate
of the contour velocity.

By increasing $f$ even further we see that the non--monotonicity
disappears (the resisting force is no more needed to unfold the
protein), the translocation speed saturates to $1/N$, meaning that the
unfolding rate is so large that in most cases unfolding is immediate,
and after that an aminoacid is imported at each time step. In this
regime, the critical value of $R$ above which translocation is
forbidden tends to $f$.

In the above results, forces and times have been reported in arbitrary
units. The simplicity of the model is its strength, making it exactly
solvable, but does not allow a full quantitative correspondence with
the experimental results. At the order of magnitude level, we can say,
by comparing our translocation velocity at saturation (large $f$,
small $R$) with that reported in \cite{Bustamante-Cell11}, that our
time unit should be of order $10^{-2}$ s.   Indeed,
  according to our setting, about $N/\langle T\rangle=1$ aminoacid per
  unit of time translocate at saturation, and this value must be
  identified with the maximum of 80 aminoacids per second found in
  \cite{Bustamante-Cell11}. On the other hand, it is known that this
model fails to accurately predict forces quantitatively: see
\cite{IP-PRL08} for a discussion of this issue and of the need to
introduce a suitable rescaling factor, and \cite{CIP-PRE11} where it
is shown that the models can reproduce only qualitatively the
hierarchy of GFP unfolding forces, when forces are applied in
different directions. We therefore do not aim at a full quantitative
discussion of force values and remain at the order of magnitude level:
at this level, by observing that the GFP equilibrium unfolding force
was reported \cite{Rief-PNAS04} to be $\simeq 35$ pN, that the stall
force in \cite{Bustamante-Cell11} was estimated to be around $20$ pN,
and that the corresponding forces in our model are around a few units,
we can estimate that our force unit should lie somewhere in between a
few pN and 10 pN.

\begin{figure}
\includegraphics*[width=0.48\textwidth]{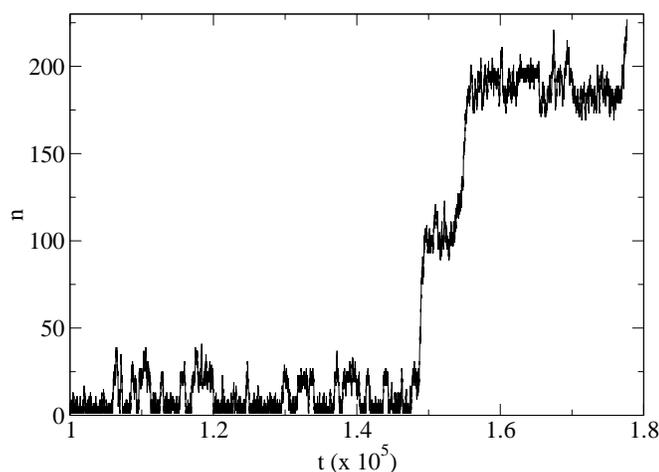}
\caption{A sample trajectory showing the intermediate. \label{fig:Traj}}
\end{figure}

In order to better understand the nature of the intermediate state, we
have simulated our stochastic process, generating single
trajectories. In Fig.\ \ref{fig:Traj} we see a sample trajectory,
obtained at $f = 4, R = 3.7$, which exhibits clearly the intermediate
state. We chose these force values since the intermediate is
short--lived, $\sim 0.18$ s in the experimental conditions
\cite{Bustamante-Cell11}, which corresponds roughly to $\sim 10$ time
steps in our model. This is indeed the order of magnitude of the
duration we observe at smaller resisting forces, and here we have
chosen $f$ and $R$ almost balancing each other in order to extend the
lifetime of the intermediate. This is a state with $n \simeq 95 \div
110$ aminoacids inside the pore, which compares well to the
intermediate with a C--terminal unfolded segment of length between 97
and 107 aminoacids observed in \cite{Bustamante-Cell11}. Structurally,
this corresponds to unfolding and importing the 5 C--terminal
$\beta$--strands, numbered 7--11, which are adjacent to each other in
the native barrel structure, arranged in the order 11--10--7--8--9.

Although we cannot compare the corresponding result with experiments,
we can easily check that if one reverses the protein, by applying $f$
to the N--terminal and $R$ to the C--terminal, a different
intermediate is observed, with $n \sim 50$ imported aminoacids,
corresponding to unfolding and importing the 3 N--terminal
$\beta$--strands, numbered 1--3 and again adjacent to each other in
the native barrel structure.


%


\section{Conclusions}

In the framework of an exactly solvable Ising--like model of protein
thermal and mechanical (un)folding, we have shown how to obtain some
exact nonequilibrium results by defining a diffusive dynamics on a
suitable free energy profile. The dynamics was then reduced to a
one--dimensional biased random walk on the corresponding reaction
coordinate, which allowed to exactly compute mean first passage times
and to easily simulate single trajectories. The approach can be
applied to different situations, by choosing appropriate reaction
coordinates, like the end--to--end length in the case of mechanical
unfolding, or the number of contacts in the case of thermal or
chemical (un)folding. Here, as an illustration, we chose the problem
of the translocation of the green fluorescent protein through a
narrow, long and neutral pore, under the action of an importing and a
resisting force. The natural reaction coordinate in this case is the
number of aminoacids imported into the pore. Our analysis focused on
two aspects of the phenomenon, which can be described at the level of
our extremely simplified model. The mean translocation time was
exactly computed as a function of the forces, showing an interesting
non--monotonic dependence on the resisting force, consistent with
recent experimental observations \cite{Bustamante-Cell11}. Simulating
single trajectories of our random walk, we have found evidences of a
two--stage unfolding phenomenon, with an intermediate state whose
number of imported aminoacids, and hence structure, are the same
observed in the experiment \cite{Bustamante-Cell11}.




\bibliographystyle{eplbib}
\bibliography{Translobib}

\begin{thebibliography}{10}
\expandafter\ifx\csname url\endcsname\relax\def\url#1{\texttt{#1}}\fi

\bibitem{ZimmBragg}
\Name{Zimm B. \and Bragg J.} \REVIEW{J. Chem. Phys.}{31}{1959}{526}.

\bibitem{PolandScheraga}
\Name{Poland D. \and Scheraga H.} \REVIEW{J. Chem. Phys.}{45}{1966}{1464}.

\bibitem{WS1}
\Name{Wako H. \and Sait\^{o} N.} \REVIEW{J. Phys. Soc. Jpn}{44}{1978}{1931}.

\bibitem{WS2}
\Name{Wako H. \and Sait\^{o} N.} \REVIEW{J. Phys. Soc. Jpn}{44}{1978}{1939}.

\bibitem{ME1}
\Name{{Mu\~noz} V., Thompson P.~A., Hofrichter J. \and Eaton W.~A.}
  \REVIEW{Nature}{390}{1997}{196}.

\bibitem{ME2}
\Name{{Mu\~noz} V., Henry E.~R., Hofrichter J. \and Eaton W.~A.} \REVIEW{Proc.
  Natl. Acad. Sci. USA}{95}{1998}{5872}.

\bibitem{Finkelstein}
\Name{Galzitskaya O. \and Finkelstein A.} \REVIEW{Proc. Natl. Acad. Sci.
  USA}{96}{1999}{11299}.

\bibitem{Baker}
\Name{Alm E. \and Baker D.} \REVIEW{Proc. Natl. Acad. Sci.
  USA}{96}{1999}{11305}.

\bibitem{ME3}
\Name{{Mu\~noz} V. \and Eaton W.~A.} \REVIEW{Proc. Natl. Acad. Sci.
  USA}{96}{1999}{11311}.

\bibitem{Maritan}
\Name{Flammini A., Banavar J. \and Maritan A.} \REVIEW{Europhys.
  Lett.}{58}{2002}{623}.

\bibitem{BP-PRL02}
\Name{Bruscolini P. \and Pelizzola A.} \REVIEW{Phys. Rev.
  Lett.}{88}{2002}{258101}.

\bibitem{MioJSTAT}
\Name{Pelizzola A.} \REVIEW{J. Stat. Mech.}{}{2005}{P11010}.

\bibitem{BPZ-PRL07}
\Name{Bruscolini P., Pelizzola A. \and Zamparo M.} \REVIEW{Phys. Rev.
  Lett.}{99}{2007}{038103}.

\bibitem{MarcoJSTAT}
\Name{Zamparo M.} \REVIEW{J. Stat. Mech.}{}{2008}{P10013}.

\bibitem{IPZ-PRL07}
\Name{Imparato A., Pelizzola A. \and Zamparo M.} \REVIEW{Phys. Rev.
  Lett.}{98}{2007}{148102}.

\bibitem{IPZ-JCP07}
\Name{Imparato A., Pelizzola A. \and Zamparo M.} \REVIEW{J. Chem.
  Phys.}{127}{2007}{145105}.

\bibitem{IP-PRL08}
\Name{Imparato A. \and Pelizzola A.} \REVIEW{Phys. Rev.
  Lett.}{100}{2008}{158104}.

\bibitem{IPZ-PRL09}
\Name{Imparato A., Pelizzola A. \and Zamparo M.} \REVIEW{Phys. Rev.
  Lett.}{103}{2009}{188102}.

\bibitem{CaraglioFnIII}
\Name{Caraglio M., Imparato A. \and Pelizzola A.} \REVIEW{J. Chem.
  Phys.}{133}{2010}{065101}.

\bibitem{CIP-PRE11}
\Name{Caraglio M., Imparato A. \and Pelizzola A.} \REVIEW{Phys. Rev.
  E}{84}{2011}{021918}.

\bibitem{Samori}
\Name{Aioanei D., Brucale M., Tessari I., Bubacco L. \and Samor\'\i B.}
  \REVIEW{Biophys. J.}{102}{2012}{342}.

\bibitem{Fernandez-Cell03}
\Name{Kenniston J.~A., Baker T.~A., Fernandez J.~M. \and Sauer R.~T.}
  \REVIEW{Cell}{114}{2003}{511}.

\bibitem{Kenniston-PNAS05}
\Name{Kenniston J.~A., Baker T.~A. \and Sauer R.~T.} \REVIEW{Proc. Natl. Acad.
  Sci. USA}{102}{2005}{1390}.

\bibitem{Sauer-NSMB08}
\Name{Martin A., Baker T.~A. \and Sauer R.~T.} \REVIEW{Nat. Struct. Mol.
  Biol.}{15}{2008}{1147}.

\bibitem{Lang-Cell11}
\Name{Aubin-Tam M.-E., Olivares A.~O., Sauer R.~T., Baker T.~A. \and Lang
  M.~J.} \REVIEW{Cell}{145}{2011}{257}.

\bibitem{Bustamante-Cell11}
\Name{Maillard R.~A., Chistol G., Sen M., Righini M., Tan J., Kaiser C.~M.,
  Hodges C., Martin A. \and Bustamante C.} \REVIEW{Cell}{145}{2011}{459}.

\bibitem{Makarov-JCP05}
\Name{Huang L., Kirmizialtin S. \and Makarov D.~E.} \REVIEW{J. Chem.
  Phys.}{123}{2005}{124903}.

\bibitem{Makarov-PSSB06}
\Name{Kirmizialtin S., Huang L. \and Makarov D.~E.} \REVIEW{Phys. Stat. Sol.
  B}{243}{2006}{2038}.

\bibitem{Cecconi-JPCB09}
\Name{Ammenti A., Cecconi F., {Marini Bettolo Marconi} U. \and Vulpiani A.}
  \REVIEW{J. Phys. Chem. B}{113}{2009}{10348}.

\bibitem{Cecconi-PM11}
\Name{Chinappi M., Cecconi F. \and Casciola C.~M.} \REVIEW{Phil.
  Mag.}{91}{2011}{2034}.

\bibitem{Stan-PNAS11}
\Name{Kravats A., Jayasinghe M. \and Stan G.} \REVIEW{Proc. Natl. Acad. Sci.
  USA}{108}{2011}{2234}.

\bibitem{Feller}
\Name{Feller W.} \Book{An Introduction to Probability Theory and Its
  Applications, vol.\ 2} (Wiley, New York) 1966.

\bibitem{Bell}
\Name{Bell G.} \REVIEW{Science}{200}{1978}{618}.

\bibitem{KumarLi}
\Name{Kumar S. \and Li M.} \REVIEW{Phys. Rep.}{486}{2010}{1}.

\bibitem{Rief-PNAS04}
\Name{Dietz H. \and Rief M.} \REVIEW{Proc. Natl. Acad. Sci.
  USA}{101}{2004}{16192}.

\bibitem{Rief-PNAS07}
\Name{Mickler M., Dima R., Dietz H., Hyeon C., Thirumalai D. \and Rief M.}
  \REVIEW{Proc. Natl. Acad. Sci. USA}{104}{2007}{20268}.

\end{thebibliography}

\end{document}